IAC–24–C1.1.8.x89332

# A causal learning approach to in-orbit inertial parameter estimation for multi-payload deployers

**Konstantinos Platanitis**[a*], **Miguel Arana-Catania**[b], **Saurabh Upadhyay**[c], **Leonard Felicetti**[d]

[a] *Faculty of Engineering and Applied Sciences, Cranfield University, United Kingdom, k.platanitis@cranfield.ac.uk*
[b] *Faculty of Engineering and Applied Sciences, Cranfield University, United Kingdom, miguel.aranacatania@cranfield.ac.uk*
[c] *Faculty of Engineering and Applied Sciences, Cranfield University, United Kingdom, saurabh.upadhyay@cranfield.ac.uk*
[d] *Faculty of Engineering and Applied Sciences, Cranfield University, United Kingdom, leonard.felicetti@cranfield.ac.uk*
[*] *Corresponding author*

**Abstract**

This paper discusses an approach to inertial parameter estimation for the case of cargo carrying spacecraft that is based on causal learning, i.e. learning from the responses of the spacecraft, under actuation. Different spacecraft configurations (inertial parameter sets) are simulated under different actuation profiles, in order to produce an optimised time-series clustering classifier that can be used to distinguish between them. The actuation is comprised of finite sequences of constant inputs that are applied in order, based on typical actuators available. By learning from the system's responses across multiple input sequences, and then applying measures of time-series similarity and F1-score, an optimal actuation sequence can be chosen either for one specific system configuration or for the overall set of possible configurations. This allows for both estimation of the inertial parameter set without any prior knowledge of state, as well as validation of transitions between different configurations after a deployment event. The optimisation of the actuation sequence is handled by a reinforcement learning model that uses the proximal policy optimisation (PPO) algorithm, by repeatedly trying different sequences and evaluating the impact on classifier performance according to a multi-objective metric.

## 1. Introduction

System identification, which directly depends on knowing the parameters that affect both dynamics and responses under actuation, is a prerequisite in all forms of control[1, 2]. The same is true in the case of spacecraft control systems, and in particular for attitude control which exhibits nonlinear, coupled dynamics [3]. After obtaining an accurate system model, classical control techniques include state observers that act upon an extended state (i.e., where the parameters are included in the state vector) such as Kalman filters in all flavours [4, 5] and even in cases of orbital capture, with two-body systems[6]. More modern approaches that include particle filtering[7, 8], predictive filter algorithms[9], as well as ML techniques[10–12] have also been studied extensively.

In the particular case of payload deploying spacecraft, the inertial parameters will inevitably change after each deployment event, given the payload mass that is ejected. This change affects both the centre of gravity as well as the inertial tensor for the spacecraft, in a way that is predictable based upon the payload's inertial properties and position it is located when onboard as depicted in Fig. 1. The motivation behind this work is to take advantage of those a priori known and expected changes, and suggest an additional approach to parameter estimation that may be used in conjunction with the currently employed methods. The aforementioned methods all require to some extent the utilisation of actuators, given the coupled nature of the system dynamics, in order to converge on the inertial parameters. The approach proposed herein is an extension of work presented in [13], which is itself an application of the techniques discussed in [14]. The suggested approach consists of two parts:

- Generation of a classifier, which employs time-series clustering (TSC) in order to learn the dynamics response of a spacecraft under actuation.

- A reinforcement learning driven model that optimised the actuation to be used, in order to guarantee classifier performance as well as optimise the actuation sequence based on multiple criteria.

The time-series clustering classifier works by studying a time-series signal, in this case to be generated through simulation. One of the benefits of the proposed method is that data need not be simulated, and is completely model-free. This allows for learning from data that includes noise, and the effects of disturbances that would be difficult to model or account for with traditional techniques. The training data source in this work will be from simulations, as discussed in Section 2, however telemetry data sources may be used where available. In this case, however, the optimisation aspect is not present as the already used actuation cannot be changed.

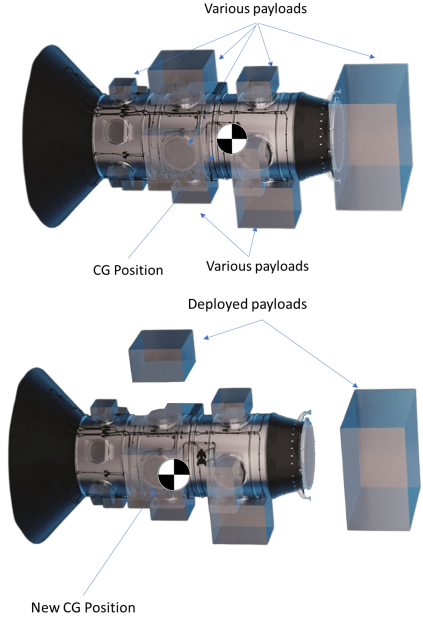

Fig. 1. Cargo carrier with payloads

## 2. Methodology

In this section, the approach taken to model a spacecraft will be discussed. The case study is for an orbital payload deployer, therefore the appropriate model generation process will be presented, as well as the modelling of actuation capabilities.

*2.1 Spacecraft dynamics*

It holds true that for any rigid object in 3D space, in an inertial frame of reference that

$$\frac{d\bar{L}_{in}}{dt} = \sum \bar{M}_{in} \quad [1]$$

Working in an inertial frame does not make sense since the inertial tensor $\mathbf{I}$, where $\bar{L} = \mathbf{I}\bar{\omega}$, will constantly change as the mass of a rotating rigid body will constantly shift. By switching to a coordinate frame that is fixed on the body and rotates with it, and substituting $\bar{L}$ we arrive at the more familiar version of Euler's equation of rigid body motion

$$\frac{d\bar{L}}{dt} + \bar{\omega} \times \bar{L} = \sum \bar{M}_i \quad [2]$$

where the torques $\bar{M}_i$ are expressed in the same body-fixed frame. Given that the angular momentum $\bar{L}$ consists of both the angular momentum of the spacecraft ($\bar{L}_{SC} = \mathbf{I}\bar{\omega}$), plus any additional stored by devices such as reaction wheels ($\bar{L}_{RW} = \sum \mathbf{I}_{RW}\bar{\omega}_{RW}$), thus Eq. (2) becomes

$$\frac{d}{dt}\left(\bar{L}_{SC} + \bar{L}_{RW}\right) + \bar{\omega} \times \left(\bar{L}_{SC} + \bar{L}_{RW}\right) = \sum \bar{M}_i$$

$$\mathbf{I}_{SC}\dot{\bar{\omega}} + \bar{\omega} \times \left(\bar{L}_{SC} + \bar{L}_{RW}\right) + \frac{d\bar{L}_{RW}}{dt} = \sum \bar{M}_i \quad [3]$$

For the simulation of the spacecraft dynamics, Eq. (3) will be numerically integrated with respect to angular rates $\bar{\omega}$ with the Runge-Kutta 4th order method. The RK4 process is selected, because it gives more accurate results compared to simple time-stepping (or Euler's method) with similar $\delta t$ values, and thus allows for greater time steps which eases the computational load when Eq. (3) is integrated for longer time periods. A custom simulator has been created, which follows along a state-space representation approach and allows for the easy application of discrete actuation vectors, as will be discussed in Sections 2.3 to 2.5.

*2.2 Multi-body systems and inertial parameters*

For any given rigid convex body, we can define a unique point known as the centre of mass (CM). The CM is at the location where a force applied on that body would generate a purely translational motion, without any angular acceleration, and is used extensively in mechanics equations and calculations. Similarly, in the case of multi-body systems, which comprise different masses $m_i$ at different positions $\bar{r}_i$ we can define a CM. This CM may or may not be contained inside any of the bodies, and the exact position of the CM is the solution of the following equation:

$$\sum m_i(\bar{r}_i - \bar{R}_{CM}) = 0 \quad [4]$$

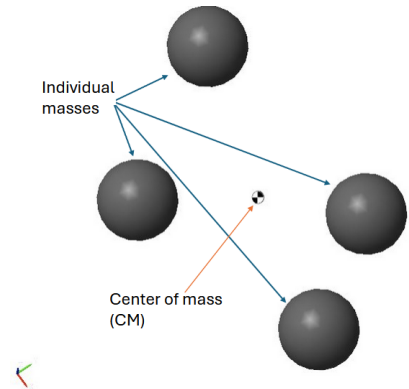

Fig. 2. Multi-body system center of mass

Having defined the location of the CM of a multi-body system, in the special case of the rotational motion of a rigid multi-body system (i.e., one where the relative positions of the individual masses do not change over time), we can define an overall inertial tensor **I** that is calculated with respect to the CM, for the system as follows:

$$\mathbf{I} = \sum \mathbf{I}_i = \sum m_i(\bar{r}_i - \bar{R}_{CM})^2 \quad [5]$$

The above approach works for point masses, whereas in this work we want to model a cargo carrying spacecraft. This means that the multi-body system will be comprised of the spacecraft itself, and all the cargo items that are onboard. In this case, the CM position can be identified as per Eq. (4) but for the derivation of the total inertia tensor a more convenient approach may be used, that of the parallel axis (or Steiner) theorem[15]:

$$\tilde{\mathbf{I}} = \mathbf{I} + \left[ \left( \bar{R} \cdot \bar{R} \right) \mathbb{1}_3 - \bar{R} \otimes \bar{R} \right] \quad [6]$$

Where $\mathbb{1}_3$ is the $3 \times 3$ identity matrix, $\bar{u} \cdot \bar{v}$ denotes the inner product of two vectors, $\bar{u} \otimes \bar{v}$ denotes the outer product of two vectors which is calculated as $(\bar{u} \otimes \bar{v})_{ij} = u_i v_j$, **I** is the inertia tensor about the body's centre of mass, and $\tilde{\mathbf{I}}$ represents the moment of inertia expressed about a point which is at a position $\bar{R}$ with respect to the CM.

If it is assumed that the inertial tensor is known for each individual piece of cargo as well as for the spacecraft itself, and that the same holds for the centre of mass position for each individual body with respect to the location of the CM of the spacecraft itself, or some other known origin. Then, an iterative procedure as follows may be used to calculate the overall system's inertial parameters:

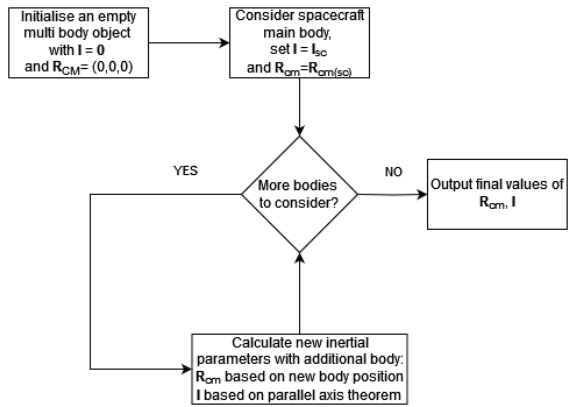

Fig. 3. Multi-body system inertial parameters calculation process

### 2.3 Actuation representation

In order to mimic actual spacecraft capabilities, the model of the spacecraft in this work will have the following types of actuators:

- Cold gas thrusters
- Reaction wheels
- Magnetorquers

All of the above actuators are able to alter the spacecraft's attitude with some being more effective than others, and with the usage of each one incurring a fuel cost, whether electrical power or fuel.

### 2.4 Gas thrusters

For the gas thrusters an approach similar to [16] will be used, where the thrusters are assumed to be of the cold gas type, and are represented in matrix form. In this format, the (column) vectors that constitute the matrix correspond with the individual torque vectors that each thruster would impart on the spacecraft, if fired. Such a matrix will inherently be of dimensions $3 \times N$ with $N$ being the number of thrusters. For a spacecraft to be controllable, a minimum of $N = 6$ thrusters is required so as to apply torques along all three axis in both directions, but it is common for a spacecraft to have multiple, redundant thruster systems and thus more actuation capabilities. A sample matrix of this representation is shown in Eq. (7) for the case of six thrusters.

$$A = \begin{bmatrix} 1 & 0 & 0 & -1 & 0 & 0 \\ 0 & 1 & 0 & 0 & -1 & 0 \\ 0 & 0 & 1 & 0 & 0 & -1 \end{bmatrix} \quad [7]$$

Given the representation of Eq. (7), the actuation of the thrusters may be considered as vector $\bar{v} \in \mathbb{R}^N$ with each individual $v_i \in [0, 1]$. The $v_i$ terms represent each thruster's actuation as a percentage of the maximum output, assuming this is to be applied with pulse width modulation (PWM). In this case, the resulting overall torque that would be applied to the spacecraft for any given is generated by $\bar{M}_t = A \cdot \bar{v}$.

The PWM actuation style means that there is a need to translate any percentage ($v_i$) between the extreme values of 0% and 100% to a signal with appropriate on/off times. This may be done with a function as follows:

$$f_{PWM}(x,t) = \begin{cases} 1, & \frac{1}{T} mod(t, T) < x \\ 0, & otherwise \end{cases} \quad [8]$$

By specifying a particular value for the frequency ($\frac{1}{T}$) of the PWM carrier signal, the function described in Eq. (8)

will generate the appropriate on/off durations based on a percentage ($x$), with respect to time. Considering the finite time required for the movement of the components in any mechanical valve, the actual response of the valves is passed through a low-pass filter, to better emulate the behaviour of the system.

*2.5 Reaction wheels*

Reaction wheels (RW) are utilised for attitude control, by taking advantage of angular momentum conservation. They consist of spinning flywheel that is driven by an electric motor, and by varying the speed of the flywheel a reaction torque is applied to the spacecraft due to conservation of angular momentum. In typical configurations, at least 3 reaction wheels are used where their axes are perpendicular to each other, in order to allow for application of torque in all directions. More than 3 modules may be present for redundancy reasons, as depicted in Fig. 4.

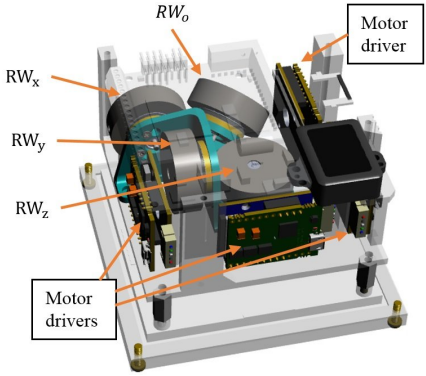

Fig. 4. Reaction wheel module arrangement [17]

For the simulation and modelling of reaction wheels, a simplified DC motor subsystem is considered as shown in Fig. 5

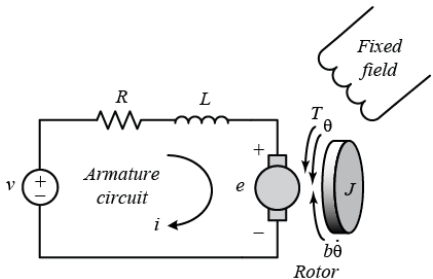

Fig. 5. Circuit representation of a DC motor

By analyzing this system, making use of Newton's and Kirchoff's laws, we can obtain the following dynamics equations:

$$J\ddot{\theta} + b\dot{\theta} = Ki$$
$$L\frac{di}{dt} + Ri = V - K\theta \quad [9]$$

where $J$ is the rotor's inertia, $\dot{\theta}$ the angular velocity, $R$ and $L$ represent the motor's resistance and inductance, $b$ is a friction coefficient and $V$ the supplied voltage. Converting this to a state-space model, we get the following:

$$\frac{d}{dt}\begin{bmatrix}\dot{\theta}\\i\end{bmatrix} = \begin{bmatrix}-\frac{b}{J} & \frac{K}{J}\\-\frac{K}{L} & -\frac{R}{L}\end{bmatrix}\begin{bmatrix}\dot{\theta}\\i\end{bmatrix} + \begin{bmatrix}0\\\frac{1}{L}\end{bmatrix}\begin{bmatrix}0\\V\end{bmatrix} \quad [10]$$

Such a model can account for the finite time response of a reaction wheel motor, and encompass for the flywheel's mass and inertia in the place of rotor inertia ($J$). Supposing a typical H-bridge configuration for the motor circuit, it is assumed that the motor can be controlled by applying any voltage $V \in [-V_s, +V_s]$ with PWM, where $V_s$ is the maximum supply voltage. An extended spacecraft system that includes Eq. (10) as a second order integrator will be used, to capture the state of each reaction wheel module.

*2.6 Time-series clustering*

Time-series clustering (TSC) is an unsupervised machine learning technique, which enables the identification of time-series datasets based on their similarity. To achieve this result, variations of the $k-$means algorithm may be used, along with an appropriate metric function. The technique may be applied to multi-dimensional signals, and depending upon the choice of metric function such a classifier may show robustness with respect to time-shifts in the dataset.[18]. An example of TSC's results on spacecraft dynamics responses with noisy sensors and actuation system variation in performance is shown in Figs. 6 and 7:

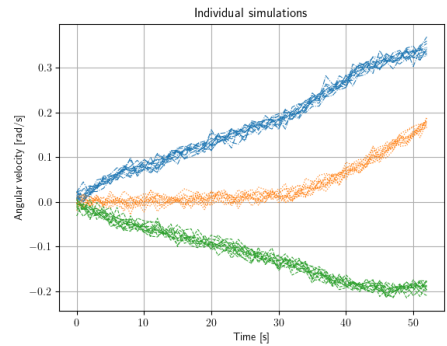

Fig. 6. Multiple simulations of dynamics responses, with noise and disturbances

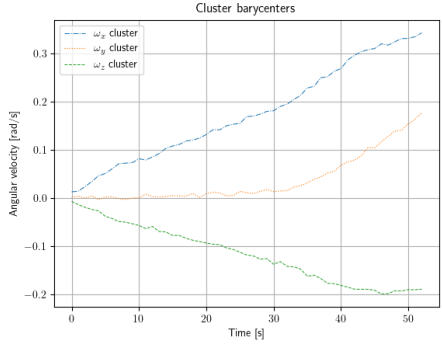

Fig. 7. TSC-identified barycenters, of the dataset shown in Fig. 6

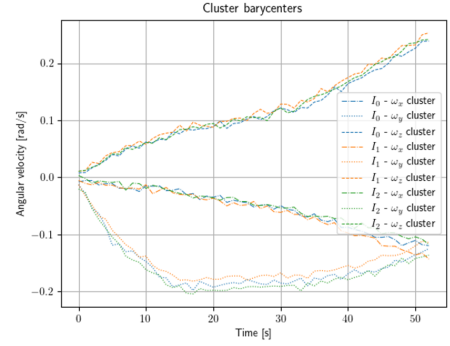

Fig. 9. TSC-identified barycenters, of the dataset shown in Fig. 8

Given the demonstrated results of TSC filtering applied to noisy dynamics, the next step is to try and apply the same technique when there are multiple datasets, originating from different system model configurations (i.e., different inertial parameters). In this case, the classifier will be able to group the responses based on their similarity and match them with the system model that generates them. The results of this process can be seen in Figs. 8 and 9.

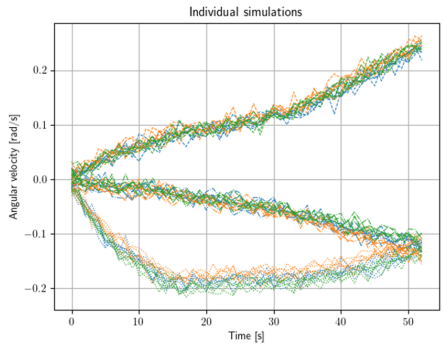

Fig. 8. Raw data from concurrent simulations of multiple inertial tensors under the same actuation

In Fig. 8 the simulation of three different inertial tensors is displayed, each with a different colour. The dataset as a whole is then given as training data to a $k$-means clustering with the SoftDTW metric, which identifies the following barycentres as seen on Fig. 9:

While the identified barycentres start with a very similar trajectory through time, after about $t = 5s$ we can observe that the $\omega_y$ barycenters start to diverge between them, followed by the barycenters of $\omega_z$ at about $t = 25s$ and finally the $\omega_x$ at about $t = 40s$.

In this work, the time-series clustering part was done with the tslearn library[*] [19], which readily implements both the appropriate $k$-means algorithm as well as the different metrics that are used in this work. To accommodate for the fact that TSC is an unsupervised learning technique, and in order to evaluate the results of the classifier versus the ground truth, after a classifier is generated the appropriate $F_1$ score is calculated for all permutations of the produced data labels. In the case of perfect classification, there will be only one permutation that produces a score of $1.0$, and this will be used as a map between actual data labels and classifier data labels.

### 2.7 Actuation sequence optimisation with reinforcement learning

Given that the resulting dynamics profiles depend upon the actuation sequence, the issue of picking such a sequence arises. Based upon the sequence chosen, the performance of the classifier will vary both when learning and when trying to identify unknown datasets, with bad choices leading to total classifier failure (i.e., generation of dynamics responses that are not discernible from one another). This problem is highly non-convex and not easy to solve analytically, and it is better evaluated by the end results of the actions that have been chosen, and not the individual steps. Working with these parameters, a reinforcement learning (RL) trained agent is a viable candidate solution to the problem.

In order to split the continuous choice into smaller, tractable pieces, the system simulator has been designed with the ability to apply any specific actuation vector for a predetermined amount of time. Such a vector would include all the necessary parameters for all the types of actuators modelled, i.e. thruster percentages (for all the thrusters) and reaction wheel supply voltages (for all the RW modules). The overall simulation is designed to conclude when for a specific number of slots in the actuation sequence, all the different actuation vectors have been applied in order. In this framework, the task of the RL agent is to optimize the selection of these actuation vectors, so as to maximise the effectiveness of the TSC algorithm in identifying the model parameters.

---

[*] `https://github.com/tslearn-team/tslearn`

For this work, the Proximal Policy Optimization (PPO) algorithm has been chosen, which is an on-policy algorithm that may be used with both discrete and continuous action spaces [20]. The action space in this case corresponds to the values of the actuation vector, and is continuous. For the implementation of the PPO algorithm, the Stable Baselines[†] library has been chosen as it already includes an implementation that supports the required options.

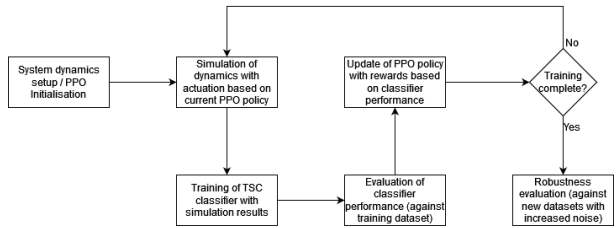

Fig. 10. Overall flowchart of the optimisation process

The observation space of the RL agent, which is used in the generation and tuning of the policy tables, is one dimensional and discrete. The observable quantity that is made available to the RL agent (i.e. the observation space) is the number of steps that the simulator has already taken. Such an approach was taken in this work, in order to completely decouple the state of the spacecraft from the actuation policy, given the end goal of generating an optimal sequence of discrete actuation vectors. In order to evaluate the results of the agent, a suitable step reward function is used, which is evaluated after each actuation application step. The RL agent thus uses this reward as feedback, in order to optimize the policy.

For the step reward function, a two-fold approach is introduced: there is a (positive) reward that is based upon the classifier's performance, and a (negative) penalty that is assigned based on a cost metric for the actuation that is chosen in each step. A constant penalty is also included for each actuation step that is taken. The reward for each step is defined as the following equation:

$$\mathbb{R}_{step} = -a_0 C_t - a_1 C_{GT} - a_2 C_{RW} + a_3 P \quad [11]$$

where $C_t$ is the cost for each timestep, $C_{GT}$ is the cost related to gas thruster usage, $C_{RW}$ is the cost related to reaction wheel usage, $P$ is the $F_1$ score, and $a_i$ are (constant) weights. The format of Eq. (11) allows for the easy adjustment of the weights on each term, and thus offers the possibility of steering the multi-objective optimization problem towards a desired trait (e.g. speed of convergence, or minimization of the used energy).

### 3. Simulation and Results

For simulation purposes, standardized bus platforms are taken into consideration such as the ARROW platform[‡] as well as other designs such as those presented in [21, 22]. Drawing from the aforementioned, a reasonable payload mass is considered to be around $m_d = 200Kg$. Given the information available for current vehicles' final stages, such as Falcon 9 second stage with a dry weight of approx $5t$ and around $100t$ of fuel on separation, assuming a GTO transfer and 5-10% of fuel remaining the spacecraft mass is assumed to be $m_{sc} = 10t$.

Based upon published dimensions of cargo and spacecraft stages, a crude approximation of the inertia tensor for each body was calculated assuming uniform mass distribution for all the bodies, which then may be used with Eqs. (5) and (6) to calculate the total moment of inertia of spacecraft and cargo.

*3.1 PPO Hyperparameters*

For the PPO algorithm, the reward function used is as described in Eq. (11) and the training hyperparameters are as shown in Table 1. All other available parameters are set to default values, as defined in the Stable Baselines library.

| Parameter | Value | Parameter | Value |
|---|---|---|---|
| n steps | 2048 | learning rate | 0.0003 |
| batch size | 64 | gamma | 0.99 |
| gae lambda | 0.95 | clip range | 0.2 |
| ent coef | 0.15 | vf coef | 0.5 |
| max grad norm | 0.5 | | |

Table 1. PPO training hyperparameters

---

[†] https://github.com/DLR-RM/stable-baselines3

[‡] https://www.airbus.com/en/products-services/space/telecom/constellations

In order to be able to use the PPO algorithm with the spacecraft simulator, the OpenAI/Farama Foundation Gymnasium library[§] was used. To this end, a custom environment was created encapsulating the numerical integrator with the time-stepping ability. The observation space of the custom environment is a singular value that represents the number of steps taken. For internal usage within the environment, the angular rates of the spacecraft are available as measured by the simulated onboard sensors, and correspond to an unbounded box that maps to $\mathbb{R}^3$. This information is not made available to the RL agent, since it is not required. The action space is of the box type, which in this case is bounded. The intervals of the dimensions that have to do with cold gas thrusters are each bounded to $[0, 1]$ whereas the intervals of the dimensions pertaining to the reaction wheels are each bounded to $[-1, 1]$. This space includes the null action, i.e. the possibility for no action to be taken by the agent. The environment is terminated if angular rates exceed predefined maximum values or the maximum number of steps is reached, and is truncated when the classifier's $F_1$ score reaches $1.0$. Next are presented several optimisation scenarios.

### 3.2 Convergence speed optimisation scenario

In order to allow for convergence speed without any restrictions, the reward function Eq. (11) is used with similar weights $a_1$ and $a_2$. This allows for the usage of all types of actuators without any preference. The terms $a_0$ and $a_3$ are adjusted to fulfil $a_0, a_3 >> a_1, a_2$. This configuration makes the RL agent try to converge as fast as possible. The reward convergence results are shown in Fig. 11. The training progress can be seen by the increase in the mean episode reward. The actuator utilisation can be seen in Fig. 12. It shows how the RL agent chooses a mix of both thrusters and reaction wheels.

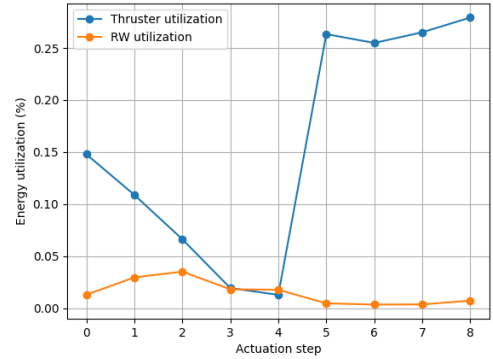

Fig. 12. Actuator utilisation

### 3.3 Fuel (gas) use optimisation scenario

In order to verify the capability of shifting the RL model choices by modifying the reward function, the $a_1$ and $a_2$ terms of the reward function Eq. (11) are adjusted so that $a_1 >> a_2$. This imbalance does steer the RL model to try and avoid actuation with thrusters, as seen in Fig. 14. However, since the overall goal is to help the classifier in correctly identifying the inertial tensors the gas thrusters are used towards the end of the actuation sequence. This is to be expected, since gas thrusters have a greater effect on the spacecraft dynamics, and thus their usage amplifies differences in the dynamics response. The mean episode reward of this run can be observed in Fig. 13. It displays an increasing trend which again shows that the RL agent is making progress as training continues.

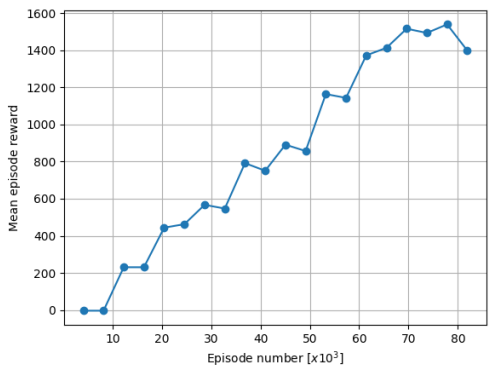

Fig. 11. Mean (episode) reward

---

§https://gymnasium.farama.org/

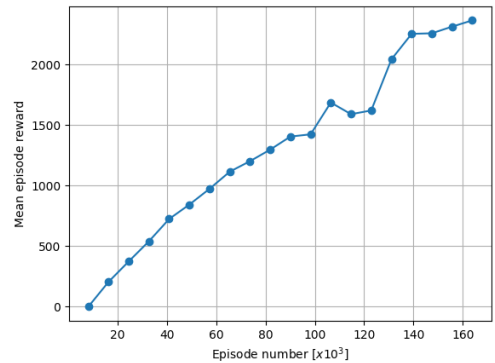

Fig. 13. Mean (episode) reward

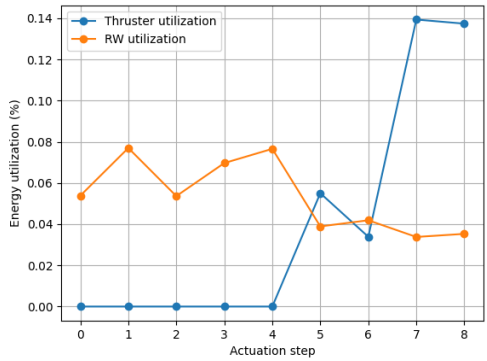

Fig. 14. Actuator utilisation

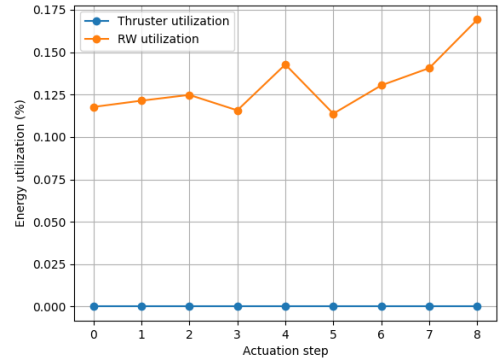

Fig. 16. Actuator utilisation

*3.4 Fuel use optimisation scenario - longer training*

As typical with RL methods, in this case more training time allows for better performance of the trained model. Without changing the parameters of Section 3.3, we allow for a much higher number of episodes (and time) during training. The pertinent results, depicted in Figs. 15 and 16 showcase that the RL model reaches a local maximum after about $150 \times 10^3$ episodes, where further training does not provide an increase in the mean episode reward. This behaviour is to be expected, given the nature of the PPO algorithm and the low entropy coefficient that has been used as per Table 1. The improvement of the model is visible in Fig. 16, by examining the minimisation of the usage of thrusters and the increase of reaction wheels utilisation.

*3.5 Robustness check*

After having trained an RL agent to pick an optimised actuation sequence, we can proceed to verify the robustness of this choice with respect to system noise. As discussed in Section 2, the simulator accounts for both sensor noise, as well as actuation noise. These noise sources are present in the datasets with which the classifier is trained, and due to the inherent robustness of TSC to noisy signals these effects are mitigated. To evaluate the level of robustness, the optimised actuation sequence is applied to simulations with increased noise levels, and the accuracy of the classifier is tested on multiple runs. The results of this analysis can be seen in Figs. 17 and 18.

As was expected, the accuracy drops off significantly with increases in noise level, and a higher degree of robustness is shown towards sensor noise than actuation (process) noise. This can be attributed to the fact that process noise induces changes to the overall dynamics response over time, whereas sensor noise can be mitigated by TSC.

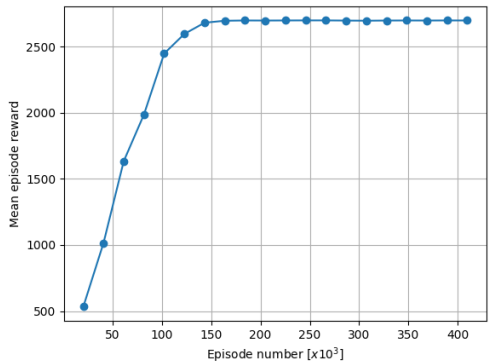

Fig. 15. Mean (episode) reward

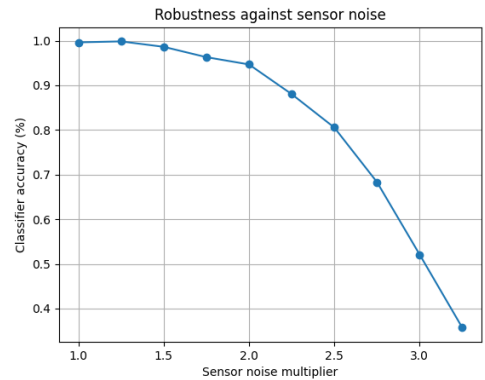

Fig. 17. Classifier accuracy vs sensor noise increase

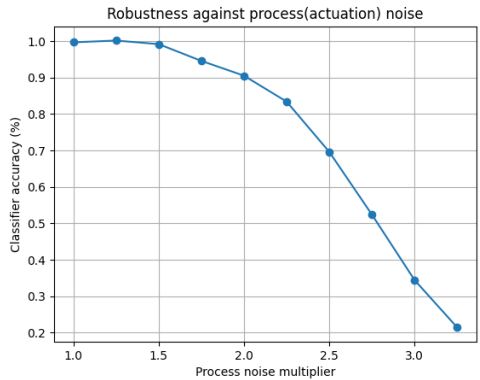

Fig. 18. Classifier accuracy vs actuation noise increase

## 4. Conclusions

As demonstrated in Section 3, the suggested approach is able to achieve the primary goal of identifying the inertial tensor of the spacecraft after payload deployment. The resulting machine learning classifiers, after training, have been tested against new datasets generated outside of the training sequence to validate robustness with respect to noise and disturbances. From the results of this analysis presented in Section 3.5, it is shown that they are robust to levels of noise up to 2 times the levels used in training. This is a direct result of training the classifier with noisy data in the first place, as well as the usage of similarity metrics for the clustering algorithm which allow for greater dissimilarities than pure Euclidian metrics.

The learning progress of the RL model which is continuously shaping the actuation profile was shown in Figs. 11, 13 and 15. As is typical in these cases, and due to the exploratory nature of the algorithm, the improvement rate is not constant and the final optimisation result depends heavily on the length of the training. However, the PPO algorithm has proven to be efficient in this particular application, with the reward function showing the capability to manipulate the end result towards a desired goal by selecting the appropriate weights for the reward objectives.

Future steps for this work include the application in different space systems, such as large structures with distributed control and sensing capabilities. Systems with more degrees of freedom for actuation, as well as additional channels of information may be easily introduced using the proposed approach. For the application on spacecraft cargo deployers, an extension of the model representation that accounts for inherent disturbances due to shifting liquid mass (fuel or cargo) is planned, for a more accurate representation of the dynamics.

**Acknowledgements**

The participation in the IAC is supported by the Centre for Autonomous and Cyber-physical Systems of Cranfield University.